\documentclass[lettersize,journal]{IEEEtran}
\usepackage{amsmath}
\usepackage{graphicx}
\usepackage{multirow}
\usepackage{subcaption}
\usepackage{booktabs}
\usepackage{todonotes}
\usepackage{comment}
\usepackage{url}
\ifCLASSINFOpdf
\else
\fi

\usepackage{xcolor}

\begin{document}
\title{When Forgetting Triggers Backdoors: A Clean Unlearning Attack}

%\author{
%	\IEEEauthorblockN{Marco Arazzi}
%	\IEEEauthorblockA{Department of Electrical, Computer and Biomedical Engineering, University of Pavia, Italy \\ marco.arazzi01@universitadipavia.it}
%	\and
%	\IEEEauthorblockN{Antonino Nocera}
%	\IEEEauthorblockA{Department of Electrical, Computer and Biomedical Engineering, University of Pavia, Italy \\ antonino.nocera@unipv.it}
%    \and
%	\IEEEauthorblockN{Vinod P.}
%	\IEEEauthorblockA{Department of Computer Applications, Cochin University of Science and Technology, Kerala, India \\ antonino.nocera@unipv.it}
%}

\author{Marco Arazzi,\thanks{Marco Arazzi and Antonino Nocera are with the Department of Electrical, Computer and Biomedical Engineering, University of Pavia, Italy. (email: marco.arazzi01@universitadipavia.it, antonino.nocera@unipv.it)}
\and Antonino Nocera,
\and Vinod P.\thanks{Vinod P. is with the Department of Computer Applications, Cochin University of Science and Technology, Kerala, India. (email: vinod.p@cusat.ac.in)}

%Telephone: (800) 555--1212, Fax: (888) 555--1212}
%\IEEEauthorblockA{\IEEEauthorrefmark{4}Tyrell Inc., 123 Replicant Street, Los Angeles, California 90210--4321}
}


%Telephone: (800) 555--1212, Fax: (888) 555--1212}


\maketitle

\begin{abstract}
Machine unlearning has emerged as a key component in ensuring ``Right to be Forgotten'', enabling the removal of specific data points from trained models. However, even when the unlearning is performed without poisoning the forget-set (clean unlearning), it can be exploited for stealthy attacks that existing defenses struggle to detect.
In this paper, we propose a novel {\em clean} backdoor attack that exploits both the model learning phase and the subsequent unlearning requests. Unlike traditional backdoor methods, during the first phase, our approach injects a weak, distributed malicious signal across multiple classes. The real attack is then activated and amplified by selectively unlearning {\em non-poisoned} samples.
This strategy results in a powerful and stealthy novel attack that is hard to detect or mitigate, highlighting critical vulnerabilities in current unlearning mechanisms and highlighting the need for more robust defenses.
%\textit{Clean \textbf{Un}learning-activated \textbf{Clean}-Label Backdoor Attack (UNCLEAN)}
\end{abstract}

% \fi
\IEEEpeerreviewmaketitle

\section{Introduction}

As the integration of AI-powered services and applications continues to grow in the modern IT landscape, the demand for high-quality and rich datasets to train and fine-tune machine/deep learning models is growing, as well.
%However, the practice of collecting low-granularity data by technology providers for the above objective poses significant challenges to protect users' privacy and allow them to control the information they disclose.
%For this reason, recent regulations and laws, such as the European General Data Protection Regulation (GDPR) and the California Privacy Rights Act (CPRA), just to mention a few, attempt to provide indications on how data produced can be collected, stored, and processed to control and limit possible privacy impacts.
%An important aspect concerns the necessity to provide users with means to consent to or refuse the use of their data for training machine and deep learning models and, generally, for feeding AI-empowered solutions.
%For this reason, a popular provision of GDPR and other privacy-related regulations and laws is the ``Right to be Forgotten'' (or the ``Right to Erasure'') \cite{mantelero2013eu}.
In line with recent regulations and laws, such as the European General Data Protection Regulation (GDPR) and the California Privacy Rights Act (CPRA), in the context of machine/deep learning, machine {\em unlearning} has been introduced as an efficient approach to allow compliance with data protection regulations in AI-based systems and implement the ``Right to be Forgotten''~\cite{mantelero2013eu}.
%Over the years, different approaches have been developed targeting the ``unlearning" paradigm to allow the removal of the contribution of specific datasets from an already trained model while preserving its performance and utility~\cite{gandikota2023erasing,kurmanji2023towards,jia2023model}. 
%However, because unlearning procedures cause variations in the original trained model, security threats arise~\cite{nicolazzo2025secure}. In the recent literature, some authors propose the use of machine unlearning to successfully carry out a backdoor attack \cite{liu2024backdoor,zhang2023backdoor,huang2024uba}. 
Various unlearning methods have been proposed to remove specific data contributions from trained models while preserving utility~\cite{chundawat2023can,golatkar2020eternal,chen2023boundary}. However, changes introduced by unlearning can create security vulnerabilities~\cite{nicolazzo2025secure}, and recent work shows that unlearning itself can be exploited to launch backdoor attacks~\cite{liu2024backdoor,zhang2023backdoor,huang2024uba}.
These approaches differ from traditional backdoor attacks, as they split the attack strategy into two phases: {\em (a)} the learning phase, in which the attacker hides an invisible trigger to a portion of his controlled training data and {\em (b)} the unlearning phase, in which the attacker requires an unlearning procedure on suitably crafted/chosen data points whose removal ultimately boosts the performance of the trigger injected during the previous phase and, therefore, allows the activation of the backdoor.
In this setting, we distinguish between two types of backdoor attack on machine unlearning, based on the strategy adopted in the second phase: {\em clean} \cite{liu2024backdoor} versus {\em poisoned} unlearning \cite{zhang2023backdoor,huang2024uba}.
Although the two categories share some similarity, the use of poisoning strategies during the unlearning phase limits the attack plausibility in scenarios where appropriate {\em aggressive} defenses can be deployed.
Indeed, while it may seem intuitive to improve anomaly detection in the forget set and filter out suspicious requests, defenses become significantly limited when forget requests consistently involve legitimate data.
However, inspired by the {\em clean} attack strategy proposed by \cite{liu2024backdoor}, we believe that no defense can actually fully prevent an attack that does not poison the forget-set during unlearning.
The approach of \cite{liu2024backdoor}, although very effective in some cases, attempts to carry out a weak and stealthy backdoor attack on one target class during training, which remains under detectability thresholds of known defenses. Then, during the second phase, it strives to amplify the attack success rate by suitably performing unlearning request of clean samples of the same target class.
According to our point of view, the idea of performing a standard backdoor attack, although weakened to remain unnoticeable, is limiting and prevents obtaining a fully powerful global attack.
In this paper, we hence propose a novel combined attack exploiting the two phases mentioned before and maintaining a fully {\em clean} unlearning phase. The main idea behind our proposal is that we do not actually need to carry out a target backdoor attack during the training phase, but it is sufficient to hide an undetectable malicious signal spread across multiple classes, rather than only the target one.
In this way, we can boost the performance of the final attack, making it powerful, stealthy, and very hard to block.
The main contribution of our approach are hence as follows.
\begin{itemize}
\item We propose ``UNlearning-activated CLEAN backdoor
attack'' (UNCLEAN), a powerful novel combined attack against machine unlearning, exploiting both the learning and unlearning phases.
\item We extend the existing literature by both exploiting only {\em unaltered} samples during the unlearning phase, thus maintaining this phase clean, and by adopting an non-targeted malicious noise injection to broader the capacity of our attack.
\item We test the performance of our attack with the main existing unlearning strategies and model architectures. We prove its robustness against existing defenses and, finally, we show the performance advantages with respect to related existing approaches.
\end{itemize}

The experimental results show the effectiveness of the proposed attack and its severity level in the context of machine unlearning. Moreover, they demonstrate the superiority of our attack over existing methods, achieving an improvement in the attack success rate of more than $32\%$ compared to the previous work by \cite{liu2024backdoor}.

\section{Related Works}

%\subsection{Machine Unlearning}
Early methods for machine unlearning primarily focus on altering training labels or reversing the learning process. Random Label unlearning~\cite{hayase2020selective} removes the influence of specific data by randomly reassigning labels, causing the model to unlearn meaningful associations with the targeted samples. Similarly, Gradient Ascent unlearning~\cite{golatkar2020eternal} takes the opposite approach of standard training by maximizing the loss for the forgotten data, effectively erasing its impact. While both techniques can achieve unlearning, they often require careful tuning to prevent performance degradation and still involve some degree of retraining.  
To enhance efficiency, Fisher Forgetting~\cite{golatkar2020eternal} modifies model weights to remove the influence of specific data points. This technique does not require access to the original training data but instead relies on the model's learned weights and the Fisher Information matrix to adjust parameters in a way that effectively removes the targeted data's impact.
More advanced strategies manipulate the model’s decision boundaries directly. Boundary Unlearning~\cite{chen2023boundary} provides a fast and effective approach to forgetting an entire class from a trained deep neural network. Rather than retraining the model from scratch, this method strategically adjusts decision boundaries so that the model no longer retains knowledge of the forgotten class while preserving its accuracy on the remaining data.  
One of the most sophisticated approach is Bad Teacher unlearning~\cite{chundawat2023can}, which employs a student-teacher framework to facilitate unlearning without requiring full retraining. In this method, a student model is trained using both competent and incompetent teachers. The competent teacher provides accurate knowledge, while the incompetent teacher introduces deliberate inaccuracies, guiding the student model to forget specific information.

%\subsection{Backdoor Attacks}
Backdoor attacks represent a class of adversarial strategies in which an adversary injects a malicious trigger into the training process to implant hidden functionality within the model. During inference, the model preserves standard behavior on clean, unaltered inputs, but misclassifies inputs containing the trigger in a way predefined by the attacker. Such attacks are effective in both centralized and distributed learning paradigms~\cite{bagdasaryan2020backdoor,xu2022more,arazzi2023turning,arazzi2024let}.
BadNet~\cite{gu2019badnets}, has been explored in recent research to highlight security risks associated with outsourcing model training. In a BadNet attack, a model is compromised during training, either by an untrusted third-party service or through transfer learning using a pre-trained model. This attack differs from adversarial perturbations, as it embeds the trigger within the model itself, rather than altering the input data.
Blended~\cite{chen2017targeted, li2021invisible} backdoor poisoning attacks aim to remain stealthy by injecting only a few poisoned samples with subtle backdoor triggers that are difficult to detect. These attacks are often studied under weak threat models, where the attacker has no knowledge of the victim model or training data. Two key strategies dominate this space: (i) using a single input instance as a universal key, and (ii) employing a trigger pattern to synthesize multiple poisoned instances. We adopt a similar stealth-oriented approach, aiming to hide the presence of the backdoor, thereby aligning with these covert poisoning strategies. 
Clean-label~\cite{zeng2023narcissus, lederer2023silent, huynh2024combat} backdoor attacks are particularly stealthy, as the poisoned data retain their original labels, making detection challenging. Notably, research has demonstrated that such attacks can be executed with minimal knowledge of the training data, using as little as 0.5\% poisoned samples to effectively implant a backdoor. Similarly, we employ clean-label backdoor techniques to covertly implant triggers into the training process, thereby aligning with stealth-focused attack paradigms and reinforcing the critical security threats such attacks pose to machine learning systems.

%\subsection{Backdoor on Unlearning}
Machine unlearning has become a pivotal mechanism for eliminating specific data from trained models to uphold user privacy and ensure regulatory compliance. However, this capability also introduces novel security vulnerabilities—most notably, backdoor attacks that exploit the unlearning process itself to subvert model integrity. As demonstrated in~\cite{liu2024backdoor}, an adversary can activate a backdoor in the unlearned model by merely requesting the deletion of a small portion of their previously contributed training data. This attack bypasses traditional poisoning techniques and achieves malicious influence without modifying any initial training samples, thereby presenting a covert and potent threat to unlearning-based systems. 
In~\cite{zhang2023backdoor}, the authors present a novel black-box backdoor attack that exploits machine unlearning to activate malicious behavior. The attacker injects both poison and mitigation samples into the training dataset to construct a model that appears benign. Subsequently, by issuing unlearning requests targeting the mitigation samples, the attacker incrementally removes their neutralizing effect, thereby triggering the latent backdoor.

Similarly, Huang et al.~\cite{huang2024uba} propose UBA-Inf, a backdoor attack that leverages influence-driven camouflage within the model, activated through unlearning operations. Unlike conventional backdoor strategies, UBA-Inf embeds the trigger in a highly stealthy manner, enabling fine-grained control over backdoor activation. The attacker accomplishes this by selectively unlearning camouflaged samples, which not only conceals the backdoor during standard operation but also mitigates challenges such as premature exposure and backdoor vanishing. This method significantly improves the persistence, stealth, and effectiveness of backdoor attacks in the presence of unlearning mechanisms.

The approach presented in this paper builds on previous methods but sets a more ambitious goal: maintaining the backdoor's stealth during training and activating it through clean data unlearning, without relying on camouflage or mitigation. This effectively proposes a ``clean" unlearning backdoor attack.

\section{Methodology}
\label{sec:methodology}

This section details the methodology employed to execute our proposed attack. 
\begin{comment}
We begin with the necessary preliminaries in Section~\ref{sec:preliminaries}, which outline the foundational concepts, the threat model, and the adversary’s objectives. Subsequently, in Section~\ref{sec:unclean}, we introduce our backdoor-enabled unlearning attack, \textit{UNCLEAN}, and describe its execution across three distinct phases.
\end{comment}

\begin{figure}[!ht]
    \centering
    \includegraphics[width=0.47\textwidth]{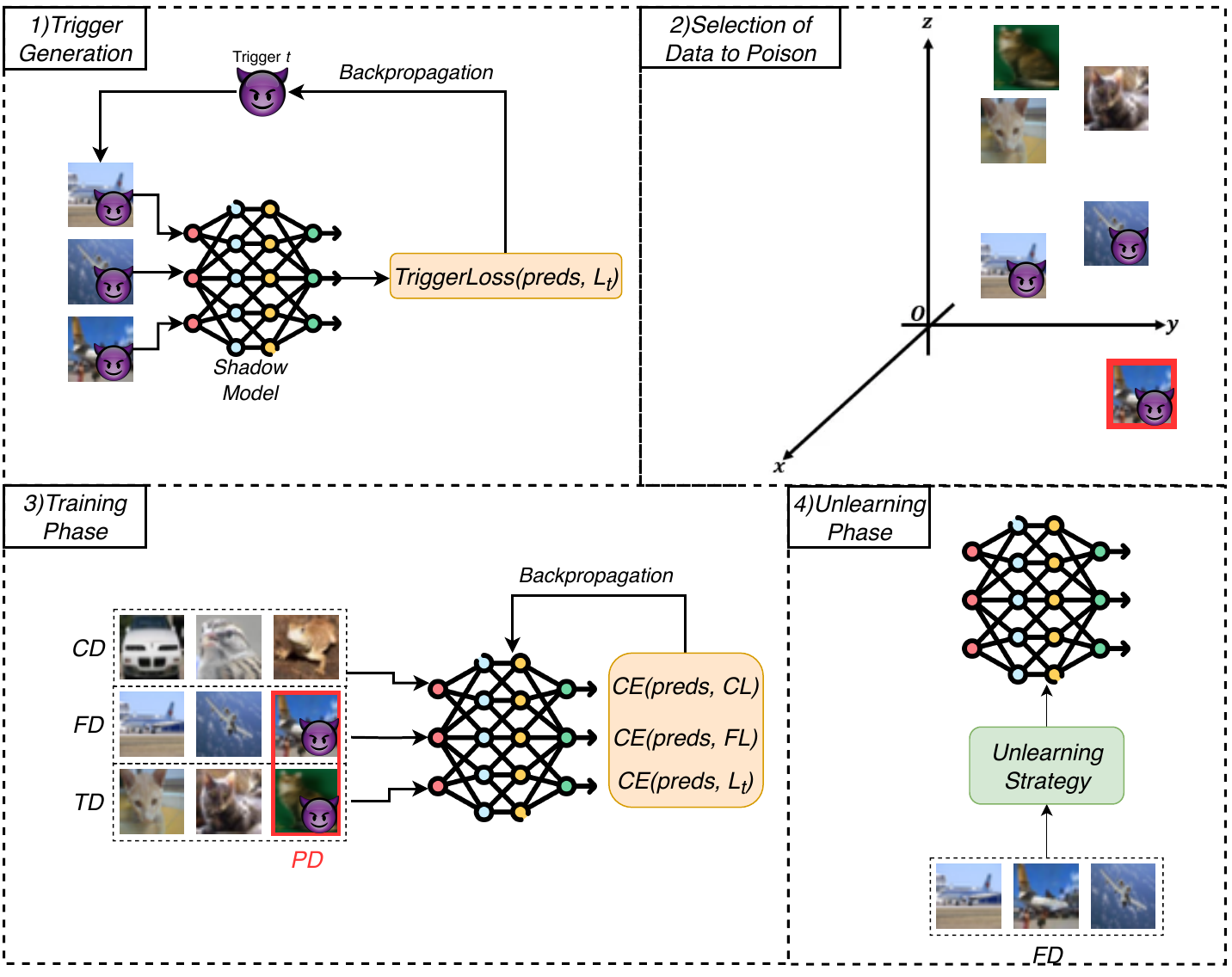}
    \caption{Pipeline (the trigger depicted in the figure serves as an example since the actual trigger is not visible). 1) Frequency Trigger generation process described in Section~\ref{sub:triggerGeneration}. 2) Selection of the best candidate data to poison in the forget-set as described in Section~\ref{sec:selectionStrategy}. 3) Training the model using the poisoned set, 4) Unlearning the forget set using only clean data.}
    \label{fig:attackPipeline}
\end{figure}

\subsection{Preliminaries}
\label{sec:preliminaries}
\subsubsection{Threat Model}

In this paper, we focus on an unlearning-activated backdoor attack that shows its full potential after the forgetting process limiting, intentionally, its performance after the initial training of the target model. 
Specifically, we assume that the attacker has access to the clean training dataset~($CD$) used to train the target model~($M$). Under this assumption, the attacker constructs a poisoned dataset~$PD = poison\_function(CD, t)$ by injecting a predefined trigger~$t$ into samples from~$CD$, while preserving their original clean labels~($CL$). This clean-label poisoning strategy embeds a backdoor into~$M$ without modifying label semantics, thereby maintaining the model’s performance on benign inputs and enhancing the stealth of the attack.

\begin{gather}
    preds = M(PD); M \leftarrow loss\_function(preds, CL)
\end{gather}

The attacker initiates the unlearning process by submitting a forget set ($FD$), which contains samples from any class except the target class. The model~$M$ undergoes unlearning to remove the influence of $FD$, resulting in an updated model~$M'$. This unlearned model~$M'= unlearn\_strategy(M, FD)$ retains high accuracy on the retain set ($RD$), which includes all non-forgotten data, but it intentionally loses the ability to correctly classify samples from $FD$.

\subsubsection{Attacker's Goal}

Targeted backdoor attacks generally focus on achieving a high rate of success against a specific target class using triggered data, while maintaining the model's accuracy on unaltered data. In this context, the attacker's goal is to embed a backdoor in the model by adding a trigger during the initial training phase, while keeping the original labels unchanged. The attack is specifically designed to underperform during training and fully manifest its effectiveness after the unlearning process. To enhance the attack's stealthiness, the forget-set provided by the attacker should consist of clean data to evade detection by filtering defenses at this stage and only need to bypass them during the training phase.

\subsection{Unlearning-activated Clean backdoor attack}
\label{sec:unclean}

%\todo[inline]{We need point-to-point differences between our proposal and \cite{liu2024backdoor,zhang2023backdoor,huang2024uba}. Let's add comments here when proposing our approaches. Explain differences in Trigger Generation and Forget-set selection...show later the advantages (perhaps in the experiments or even intuitively here if it is intuitive). Defese Cognitive Distillation (CD)~\cite{huangdistilling}} 
The core concept of this study is based on launching an unlearning-activated backdoor attack that avoids tampering with the unlearning phase (i.e., the forget-set $FD$) while only introducing poison to the initial dataset utilized for training. 
With this, we want to consider the most realistic scenario in which the attacker controls just one of the sources and has access only to partial data $TD\in CD$ from the target class $TC$ and a subset of data from the remaining classes from which the forget-set $FD$ can be selected.

Our attack consists of three main phases: trigger generation, selection of data to poison, and the training/unlearning phase (see Figure~\ref{fig:attackPipeline}).

\subsubsection{Trigger Generation}
\label{sub:triggerGeneration}

Although the backdoor can be activated using clean-label inputs, the attacker must poison the training data, making the attack difficult to detect with existing defenses or manual inspection. To enhance stealth, we generate the trigger~$t$ in the frequency domain with the same shape of the image, using the Discrete Cosine Transform (DCT), which reduces visual artifacts and preserves the natural appearance of the image. Operating in the frequency domain allows us to embed the trigger by modifying selected mid-frequency components, which are less perceptible to the human eye yet influential to neural network activations. Prior work has demonstrated that frequency-based perturbations, particularly those applied via DCT, can effectively produce stealthy triggers while remaining robust under common image transformations like compression and resizing~\cite{yu2023backdoor, wang2022invisible}.

Specifically, we embed the trigger~$t$ into an image~$I$ by modifying selected frequency components using the Discrete Cosine Transform (DCT). We begin by transforming both the image~$I$ and the trigger~$t$ into the frequency domain: 
$F_I = DCT(I), \ F_t = DCT(t)$.

Next, we generate a frequency mask $Mask(u, v)$ to constrain modifications to a specific frequency band $(f_{\text{min}}, f_{\text{max}})$, ensuring that only the desired frequency components are altered. Here, $u$ and $v$ denote the frequency indices corresponding to the horizontal and vertical components, respectively. The mask $Mask(u, v)$ is defined as follows:
\begin{gather}
    Mask(u,v) =
        \begin{cases}
            1, & f_{\min} \leq \sqrt{u^2 + v^2} \leq f_{\max} \\
            0, & \text{otherwise}
        \end{cases}
\end{gather}

We constrain the trigger to a specific frequency band $(f_{\text{min}}, f_{\text{max}})$ to limit modifications to perceptually inconspicuous components, thereby enhancing stealthiness. Additionally, we introduce a learnable parameter $\alpha$, bounded by a sigmoid function, to control the trigger’s strength in a smooth and differentiable manner. This approach allows adaptive, low-magnitude perturbations that effectively embed the backdoor while minimizing visual and statistical detectability. The learnable parameter can be obtained as: $\alpha = \sigma(\theta)$, where $\alpha \in (0,1)$, $\sigma$ is the sigmoid function and $\theta$ represents the trainable parameters.
Using the mask generated before the trigger is applied to the image as follows: %\textcolor{red}{Vinod: Avoid using $M$ in the following equation, as we also use the symbol to represent model $M$. Do we have the trigger and image of the same dimension? I think, we must clarify.}

\begin{gather}
    F_{poisoned} = F_I + \alpha\cdot M \cdot (F_t - F_I),
\end{gather}

this ensures that only the selected frequencies are affected.
The obtained $F_{poisoned}$ is then reconverted in the spatial domain using the \emph{Inverse DCT} ($IDCT$): $I' = IDCT(F_{poisoned})$.

We clamp the reconstructed image \( I' \in PD \) to the range \([0, 1]\) to maintain visual consistency. To further enhance the stealthiness of the attack, we apply regularization to both the blending coefficient \( \alpha \) and the trigger \( t \), thereby constraining the perturbations introduced to the image. In particular, $R_{\alpha} = \lambda_{\alpha} |\alpha|$ and $R_t = \lambda_t || t ||_2$, where $\lambda_{\alpha}$ and $\lambda_t$ are the regularization coefficients.

To modify the trigger, the attacker uses it on the images within the forget-set $FD$ and inputs them into a shadow model $SM$, which is trained on the attacker's data. A cross-entropy loss $CE$ is then used to compare the prediction on $I'$ with the target label $L_t$, as follow:

\begin{gather}
    pred' = SM(PD); \ t \leftarrow CE(pred', L_t).
\end{gather}

\begin{figure}[!ht]
    \centering
    \includegraphics[width=0.60\columnwidth]{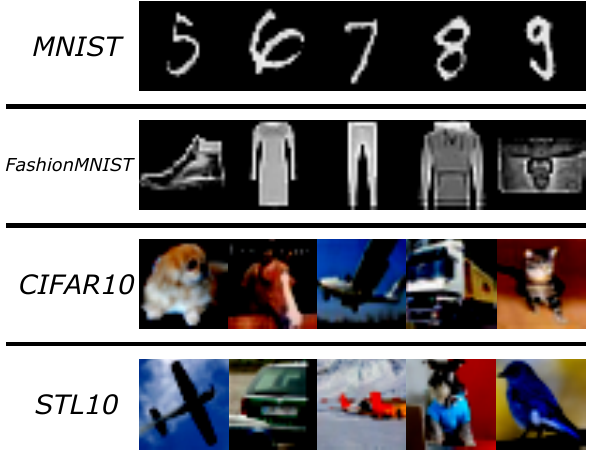}
    \caption{Examples of triggered images.}
    \label{fig:triggerExamples}
\end{figure}

Backpropagating the result of $CE$ loss function into $t$ we obtain a trigger that contain information of the target calls while preserving its stealthiness operating in the frequency domain making the trigger almost invisible. Examples of triggered images can be seen in Figure~\ref{fig:triggerExamples}.

\subsubsection{Selection of data for poisoning}
\label{sec:selectionStrategy}

As explained earlier, the attacker wants the backdoor to work only after the unlearning happens. To do this, the attacker secretly adds the trigger to some images from both the target class and the forget-set~($FD$). This spreads out the trigger’s features across different classes without changing the actual labels, making it less noticeable during training. Later, when the model is told to unlearn $FD$, which contains clean (non-triggered) versions of the poisoned images, it accidentally learns to classify triggered images as the target class. The goal is to make the backdoor effective after unlearning, without hurting the model’s performance on the remaining clean data.

\par To make the attack stronger, the attacker should choose the most effective forget-set~($FD$). We suggest picking images that, once triggered, produce internal features (latent representations) that are very different from the average features of the target class. To find this average, we use the same shadow model~($SM$) mentioned earlier.
\begin{gather}
   Avg\_TD\_embed = mean(SM.embed(TD))
\end{gather}

in the same way we generate the embeddings for the entire forget-set $FD$ and we compare the with average representation of the target class obtained before using \emph{cosine similarity} $CS$:

\begin{gather}
    FD\_embeds = SM.embed(FD); \\
    S = CS(FD\_embeds, Avg\_TD\_embed)
\end{gather}

We denote by $S$ the set of similarity scores between each element in the forget-set and the average embedding of the target class, \( \textit{Avg\_TD\_embed} \). Using these scores, the attacker identifies the indices of the samples that are least similar to the target class and subsequently selects a small subset of them for the attack.

 % where $S$ represents the set of similarity scores for each element within the forget-set relative to $Avg\_TD\_embed$. Utilizing $S$, we identify the indices of the data that exhibit the lowest resemblance to the target class and, subsequently the attacker a small percentage of these elements.

\begin{equation}
    poison\_indxs = sort(S).last(percentage)
\end{equation}

The value of \( \text{percentage} \) directly influences the strength of the backdoor attack prior to unlearning, as it determines how many low-similarity samples from the forget-set are selected and embedded with the trigger.

\subsubsection{Training/Unlearning phase}

After selecting and poisoning the data in the forget-set~($FD$), we randomly choose a subset of data ($random\_indxs$) from the target class data~($TD$) to be poisoned. This step strengthens the connection between the clean features of the forget-set and the trigger, ensuring that the signal is blended effectively. By aligning the trigger with the clean features of the target class, we facilitate a more seamless activation of the backdoor without needing further alterations during the unlearning process. This process shifts the classification towards the target class, as explained in Section~\ref{sub:triggerGeneration}. Instead of purely achieving catastrophic forgetting, we aim to facilitate gradient realignment~\cite{toneva2018an} and selective forgetting.

The poisoned data~($PD$) are then used to train the base model $M$ without altering the process other than poisoning the data:

\begin{gather}
    PD = \{FD[poison\_indxs], \ TD[random\_indxs]\} 
\end{gather}
\begin{align}
    M \leftarrow & \ CE(M(CD), CL) +  \\ \nonumber
                 & \ CE(M(FD[\textit{poison\_indxs}]), FL) + \\ \nonumber
                 & \ CE(M(TD[\textit{random\_indxs}]), L_t)
\end{align}

where, as previously mentioned, $CL$ represents the clean labels linked to the clean dataset $CD$, $FL$ denotes the clean labels that are associated with the poisoned data within the forget-set $FD$, $CE$ is the cross-entropy loss and $L_t$ refers to the target class.

Initially, the trigger $t$ is embedded in both the target class data and part of the forget-set preserving their clean labels to create a conflicting learning signal during training. Since the trigger appears across multiple classes the model learns a weaker assocition between the trigger $t$ and $L_t$, and the its decision boundary is influenced by conflicting features from $FD[poison\_indxs]$.
In our approach, we leverage machine unlearning to selectively forget specific features associated with the clean forget-set ($FD$). By applying the unlearning strategy $M' = \text{unlearn\_strategy}(M, FD)$, we realign the model's gradients, adjusting the decision boundary to eliminate the association between the target $t$ and the forget label $FL$, while preserving the backdoor association with the target label $L_t$. This selective forgetting ensures that the model forgets only the clean features of $FD$ without incorporating any new backdoor information, thereby maintaining the association between $t$ and $L_t$ while removing the mapping between $t$ and $FL$.

\section{Experimental Results}
This section presents the experimental campaign conducted to demonstrate the effectiveness of the methodology introduced in Section~\ref{sec:methodology}.

\subsection{Datasets, Models and Evaluation Metrics}
\label{sec:datasetsModels}

We conduct experiments on four benchmark datasets: MNIST~\cite{deng2012mnist}, FashionMNIST~\cite{xiao2017fashion}, CIFAR-10~\cite{krizhevsky2009learning}, and STL-10~\cite{coates2011analysis}. MNIST consists of 28×28 grayscale images of handwritten digits from 0 to 9. FashionMNIST follows the same format but represents 10 categories of clothing items. CIFAR-10 comprises 32×32 color images across 10 classes, including objects such as airplanes, automobiles, and animals. STL-10, while also covering 10 object classes similar to CIFAR-10, contains higher-resolution color images of size 96×96.

The models used instead are: ResNet18~\cite{he2016deep}, ResNet50~\cite{he2016deep}, VGG11~\cite{simonyan2015very}, AllCNN~\cite{springenberg2015striving} and ViT~\cite{dosovitskiy2020image}. 

To evaluate the effectiveness of our attack, we consider several key factors. First, it is essential to quantify the impact of the backdoor both before and after the unlearning phase, ensuring that the attack remains inconspicuous during training and becomes effective only afterward. Equally important is the requirement that the model maintains its utility throughout unlearning, specifically preserving clean accuracy on both the retained and forgotten samples. To do this, we introduce the following three evaluation metrics.

\textbf{Acc retain}: This metric measures the accuracy of the model in the retention set after the unlearning process. Quantifies whether the model successfully preserves performance in retained data, thus ensuring that unlearning does not degrade utility in unaffected information.

\textbf{Acc Forget}: Evaluate the model's performance both before and after the unlearning process on the forget-set to determine whether the targeted information has indeed been erased.

\textbf{Attack Success Rate (ASR)}: This metric quantifies the proportion of poisoned inputs that the model misclassifies as the attacker's target class. It is evaluated both before and after the unlearning phase to assess the activation and persistence of the backdoor.

%%% Vinod from here%%%
\subsection{Unlearning Strategies}
\label{sec:unlearningStrategy}

To evaluate our method, we explored various unlearning strategies to assess the generalizability of our attack and to demonstrate the points made in the methodology section across different configurations and approaches. Specifically, we chose:

\textbf{Bad Teacher Unlearning~\cite{chundawat2023can}}: This unlearning strategy leverages a student–teacher framework wherein the student model is trained using both competent and incompetent teachers to selectively forget specific data without requiring full retraining. This dual-source supervision enables the student model to attenuate the influence of targeted data.

\textbf{Fisher Forgetting~\cite{golatkar2020eternal}}: This unlearning technique introduces a scrubbing method that removes information about specific training data from a model's weights. This approach eliminates the need for full retraining or access to the original training data.

\textbf{Boundary Unlearning~\cite{chen2023boundary}}: This method provides a fast and effective mechanism for removing an entire class from a trained deep neural network (DNN) without requiring full retraining. The approach operates by strategically altering the model’s decision boundary to emulate the behavior of a reference model retrained from scratch without the target class.performance on the remaining data.

\textbf{Gradient Ascent~\cite{golatkar2020eternal}}: It is a technique for removing specific information from a trained deep neural network by reversing the learning process. Unlike standard training, which uses gradient descent to minimize a loss function, gradient ascent unlearning maximizes the loss for the targeted data, effectively erasing its influence from the model.

\textbf{Random Label Unlearning~\cite{hayase2020selective}}: This approach eliminates the influence of specific data by assigning random labels to the forget set and subsequently fine-tuning the model. Rather than directly altering model parameters, the method induces forgetting by disrupting the association between inputs and their correct labels.

\subsection{Backdoor Defenses}
\label{sec:defenses}

In the experimental campaign, we also tested the resilience of our attack against backdoor defense strategies. Since our attack is based on clean unlearning, we focused solely on defenses designed to counter backdoor attacks during the training phase.In particular, we considered the followings.

\textbf{Cognitive Distillation (CD)~\cite{huangdistilling}}: A defense method that distills a minimal pattern from input images, revealing the essential features that determine the model's output; this is used as input mask to remove redundant information.

\textbf{Neural Cleanse (NC)~\cite{wang2019neural}}: A defense method designed to detect and remove backdoor triggers in deep neural networks. It identifies potential input triggers that cause misclassifications when added to an input.

\textbf{Implicit Backdoor Adversarial Unlearning (I-BAU)}: This algorithm formulates the unlearning process as a minimax optimization problem and solves it using implicit hypergradients. 

\subsection{Experimental Setup}
\label{sec:experimentalSetup}

In the baseline experiments presented in Section~\ref{sec:baslineResults}, we employed ResNet-18 as the model architecture, optimized using stochastic gradient descent (SGD) with a learning rate of 0.1. Approximately 5\% of the training data, including a subset of the forget-set and images from the target class, was poisoned during training. All reported results are averaged over multiple runs to ensure statistical robustness. The regularization coefficient for trigger generation, $\lambda_t$, was fixed at 0.05 across all datasets, while $\lambda_{\alpha}$ varied across \{0.01, 0.001, 0.0001\}, depending on the dataset.

We evaluated the impact of data selection by choosing a forget set samples with high or low similarity to the target class and measuring the resulting attack success rate (ASR), keeping all other conditions constant. Additionally, we analyzed how ASR varied before unlearning across different poisoning rates in the forget set, ranging from 0.05 to 0.5. To assess the robustness of our attack against backdoor defenses, in Section~\ref{sec:defensesEvaluation} we tested three representative defense mechanisms introduced in Section~\ref{sec:defenses}, each targeting different aspects of the model, such as training data or parameters. These evaluations were conducted under the same baseline settings. In Section~\ref{sec:differentModels}, we investigated the transferability of the attack across various architectures using default hyperparameters, with the exception of ResNet-50 and ViT, where we set the learning rate to 0.01. Finally, in Section~\ref{sec:abaltion}, we performed an ablation study to quantify the contribution of key components in our approach.

\subsection{Evaluation of UNCLEAN}
\label{sec:baslineResults}
 
In this section, we evaluate our attack under the baseline conditions defined in Section~\ref{sec:experimentalSetup}, across all five unlearning strategies introduced in Section~\ref{sec:unlearningStrategy}. This evaluation aims to assess the general validity of our approach by examining how different unlearning mechanisms influence the effectiveness of the attack. Additionally, we verify that the poisoning process does not interfere with the normal training or unlearning behavior on clean data, consistent with standard evaluations of backdoor attacks. The results are reported in Table~\ref{tab:mainResults}.

As expected, after the unlearning process, our attack significantly improves its effectiveness, with the Attack Success Rate (ASR) in some cases more than doubling compared to the pre-unlearning baseline. This trend suggests that certain unlearning methods, while successfully degrading accuracy on the forget set, do not fully erase adversarially useful representations, inadvertently making the model more vulnerable to attacks. 
Our results confirm the intuition outlined in the methodology section, Section~\ref{sec:methodology}, where we hypothesized that gradient realignment and selective forgetting would shape the model’s post-unlearning behavior. As described, the poisoning strategy embeds the trigger in both the forget-set and the target class data, initially creating a conflicting learning signal. This weakens association between the trigger and its intended class, as the decision boundary is influenced by overlapping feature distributions. However, when unlearning is applied, the model undergoes gradient realignment, adjusting its decision boundary to remove the conflicting association between trigger and clean labels of the forget-set while preserving the backdoor association with the target label.  

The high ASR observed under Fisher Forgetting and Boundary Unlearning supports our hypothesis that unlearning strategies which selectively target the forget-set may inadvertently preserve adversarially exploitable information. Rather than completely erasing the backdoor, these methods disrupt the decision boundary in a way that enhances adversarial exploitability. This is because selective forgetting forces the model to remove specific clean features of the forget-set without introducing new backdoor information, effectively preserving the association between the trigger and the target label while only breaking its connection to the forget-set data. As a result, adversarial attacks become even more effective post-unlearning, as the model has retained the influence of the poisoned trigger on the target class while losing its conflicting associations. This finding confirms that, in the absence of explicit mechanisms to disrupt adversarial pathways, unlearning strategies may unintentionally preserve, or even reinforce, backdoor vulnerabilities instead of mitigating them.

As mentioned earlier, our attack is designed to preserve the model's inherent properties, ensuring that the unlearning of clean data remains unaffected while specifically targeting and altering the unlearning process. To evaluate this, we also performed the unlearning strategies on a clean model and the results are reported in Table~\ref{tab:unlearningResultsNoAttack}.
As we can see, the unlearning metrics are comparable between poisoned and clean models, proving that our attack does not interfere with the general functioning of the model, even in the unlearning of clean data.

\begin{table*}[h]
\centering
\scriptsize
\caption{Baseline results across datasets and unlearning strategies.}
\label{tab:mainResults}
\renewcommand{\arraystretch}{1.1}
\resizebox{0.9\textwidth}{!}{%
\begin{tabular}{l|ccc|ccc|ccc|ccc}
\toprule
\textbf{Method} & \multicolumn{3}{c|}{\textbf{MNIST}} & \multicolumn{3}{c|}{\textbf{FashionMNIST}} & \multicolumn{3}{c|}{\textbf{CIFAR10}} & \multicolumn{3}{c}{\textbf{STL10}} \\
 & Acc Retain & Acc Forget & ASR & Acc Retain & Acc Forget & ASR & Acc Retain & Acc Forget & ASR & Acc Retain & Acc Forget & ASR \\
\midrule
Before Unlearning & 99.3\% & 99.6\% & 27.7\% & 89.3\% & 85.1\% & 26.5\% & 70.8\% & 78.8\% & 28.3\% & 39.6\% & 50.8\% & 27.4\% \\
\midrule
Bad Teacher~\cite{chundawat2023can} & 99.3\% & 0.0\% & 84.1\% & 91.3\% & 0.0\% & 67.9\% & 71.5\% & 7.0\% & 78.2\% & 40.9\% & 25.1\% & 77.4\% \\
Fisher Forgetting~\cite{golatkar2020eternal} & 99.3\% & 0.0\% & 88.2\% & 89.5\% & 0.0\% & 93.4\% & 70.2\% & 0.0\% & 92.7\% & 40.4\% & 0.0\% & 91.7\% \\
Boundary Unlearning~\cite{chen2023boundary} & 95.2\% & 5.5\% & 94.2\% & 68.2\% & 17.1\% & 89.1\% & 62.2\% & 17.6\% & 88.5\% & 36.5\% & 14.4\% & 81.9\% \\
Gradient Ascent~\cite{golatkar2020eternal} & 89.5\% & 30.0\% & 47.6\% & 72.2\% & 0.0\% & 84.3\% & 66.3\% & 4.4\% & 78.6\% & 33.7\% & 17.2\% & 84.1\% \\
Random Label~\cite{hayase2020selective} & 98.1\% & 11.3\% & 60.1\% & 86.1\% & 11.4\% & 69.1\% & 65.6\% & 14.0\% & 78.6\% & 41.1\% & 44.8\% & 59.2\% \\
\bottomrule
\end{tabular}%
}
\end{table*}

\begin{table*}[h]
\centering
\tiny
\caption{Metrics (\%) of unlearning without attack across datasets.}
\label{tab:unlearningResultsNoAttack}
\renewcommand{\arraystretch}{1.1}
\resizebox{0.80\textwidth}{!}{%
\begin{tabular}{l|cc|cc|cc|cc}
\toprule
\textbf{Method} & \multicolumn{2}{c|}{\textbf{MNIST}} & \multicolumn{2}{c|}{\textbf{FashionMNIST}} & \multicolumn{2}{c|}{\textbf{CIFAR10}} & \multicolumn{2}{c}{\textbf{STL10}} \\
 & Acc Retain & Acc Forget & Acc Retain & Acc Forget & Acc Retain & Acc Forget & Acc Retain & Acc Forget \\
\midrule
Bad Teacher~\cite{chundawat2023can} & 99.1\% & 0.0\% & 92.7\% & 1.7\% & 72.9\% & 11.5\% & 40.1\% & 35.2\% \\
Fisher Forgetting~\cite{golatkar2020eternal} & 99.3\% & 6.5\% & 93.2\% & 0.0\% & 71.5\% & 0.0\% & 44.1\% & 0.0\% \\
Boundary Unlearning~\cite{chen2023boundary} & 89.3\% & 0.0\% & 80.1\% & 0.6\% & 69.2\% & 4.2\% & 36.5\% & 14.4\% \\
Gradient Ascent~\cite{golatkar2020eternal} & 89.5\% & 30.0\% & 69.5\% & 0.0\% & 70.8\% & 39.5\% & 32.0\% & 20.7\% \\
Random Label~\cite{hayase2020selective} & 97.5\% & 0.7\% & 89.5\% & 1.6\% & 62.5\% & 8.8\% & 34.8\% & 44.5\% \\
\bottomrule
\end{tabular}%
}
\end{table*}

\begin{figure}[!ht]
    \centering
    \begin{subfigure}[b]{\columnwidth}
        \centering
        \includegraphics[width=0.75\columnwidth]{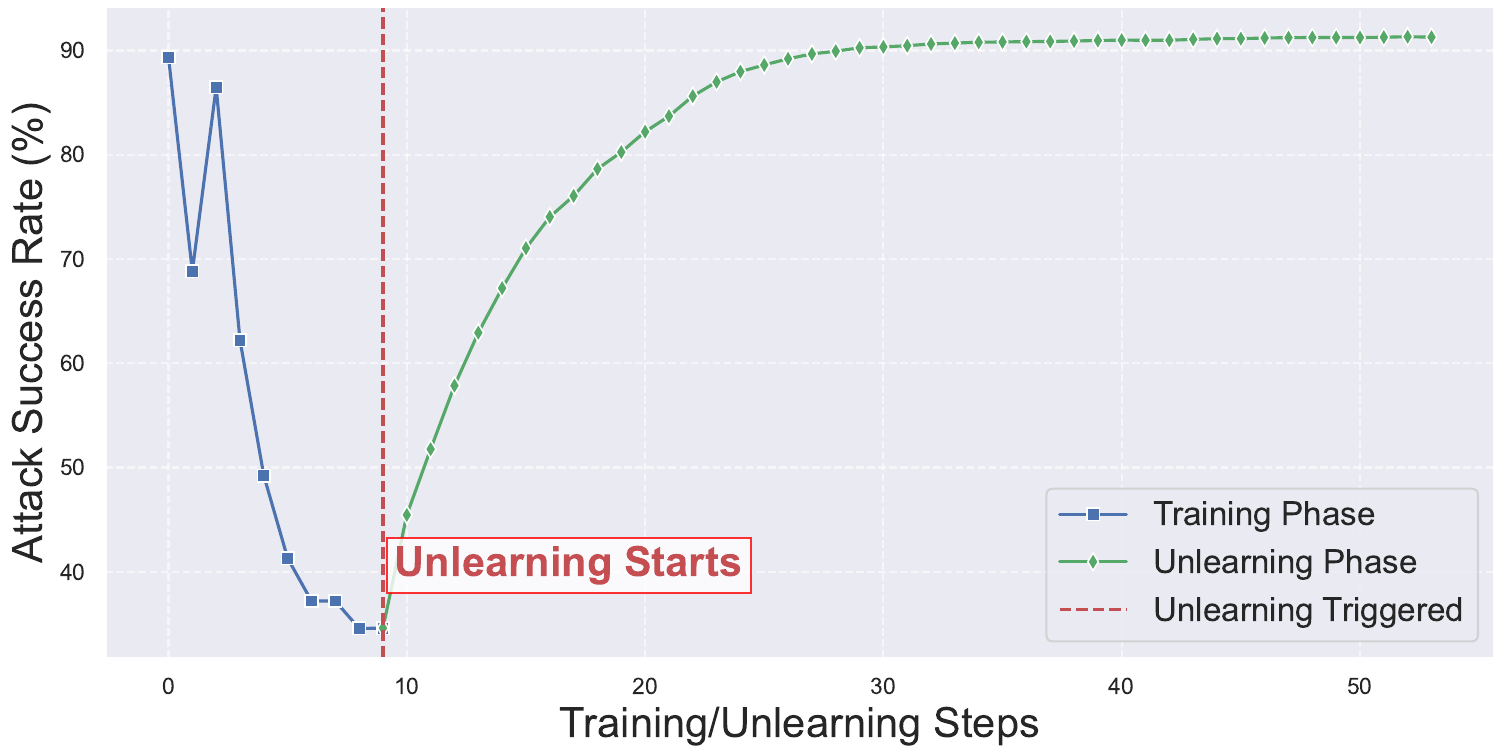}
        \caption{CIFAR10}
        \label{fig:sub1}
    \end{subfigure}
    \hfill
    \begin{subfigure}[b]{\columnwidth}
        \centering
        \includegraphics[width=0.75\columnwidth]{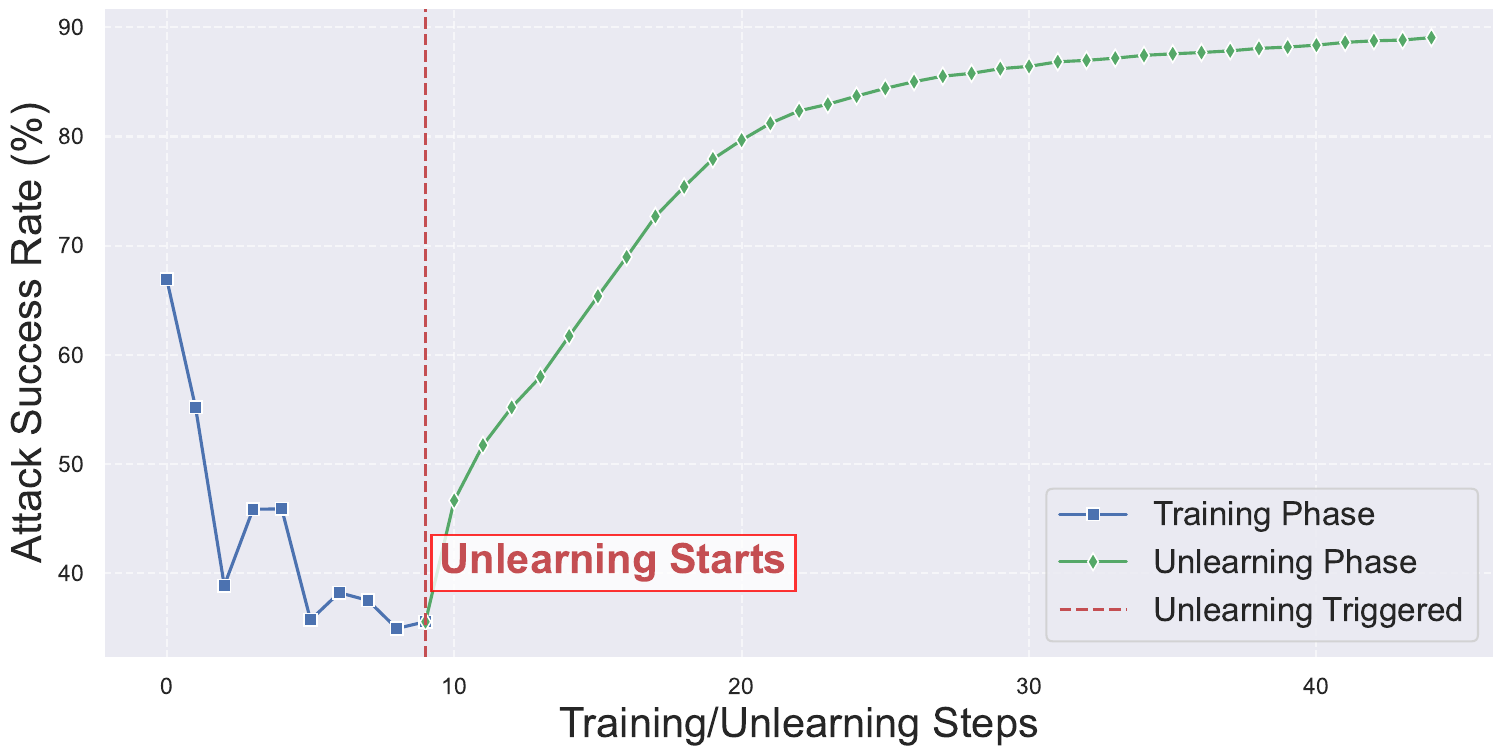}
        \caption{FashionMNIST}
        \label{fig:sub2}
    \end{subfigure}
    \caption{Evaluation stealthiness of the attack during training and activation during unlearning on CIFAR10 dataset using the Boundary Unlearning strategy.}
    \label{fig:traningUnlearningASR}
\end{figure}

Using CIFAR-10 and FashionMNIST with Boundary Unlearning as the reference strategy in Figure~\ref{fig:traningUnlearningASR}, we illustrate the evolution of the attack success rate (ASR) throughout the training and unlearning phases. The observed sharp decline in ASR during training indicates that distributing the trigger across both the target class and the forget-set effectively suppresses its influence, thereby enhancing the stealthiness of the attack prior to unlearning. In contrast, once the unlearning phase begins, the attack success rate steadily increases until convergence, revealing the full effectiveness of the backdoor.

\subsection{Evaluation of Selection Strategy}
\label{sec:selectionStrategyExp}

\begin{figure}[!ht]
    \centering
    \includegraphics[width=0.70\columnwidth]{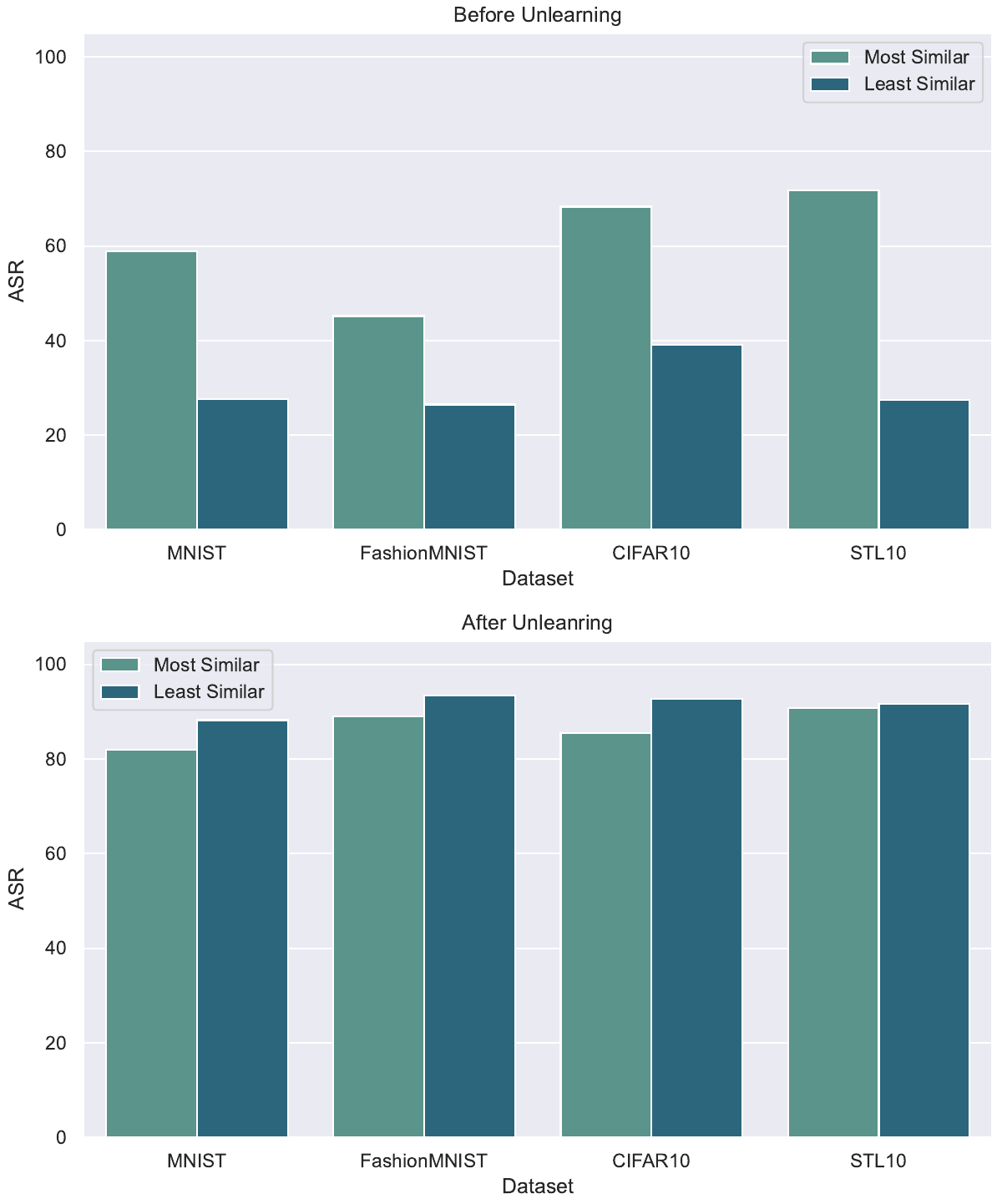}
    \caption{Evaluation of the difference between selecting the most similar poisoned data versus the most distant data relative to the target class.}
    \label{fig:distantSimilarplot}
\end{figure}

In this section, we aim to validate our hypothesis that poisoning data samples most dissimilar to the target class allows for a more effective dispersion of the trigger signal across both the target and forget sets. To test this, we compare the impact of selecting the most dissimilar versus the most similar samples within the forget set. The results, summarized in Figure~\ref{fig:distantSimilarplot}, are presented using Fisher Forgetting as the reference unlearning strategy.

As observed, the results confirm our intuition; however, when evaluating ASR after the unlearning process, both selection strategies yield comparable results, with a slight advantage observed for the distant data configuration. This suggests that using highly similar samples may reduce the overall effectiveness of the attack, as unlearning tends to partially remove information associated with both the trigger and the target class itself.

\begin{figure}[!ht]
    \centering
    \includegraphics[width=0.75\columnwidth]{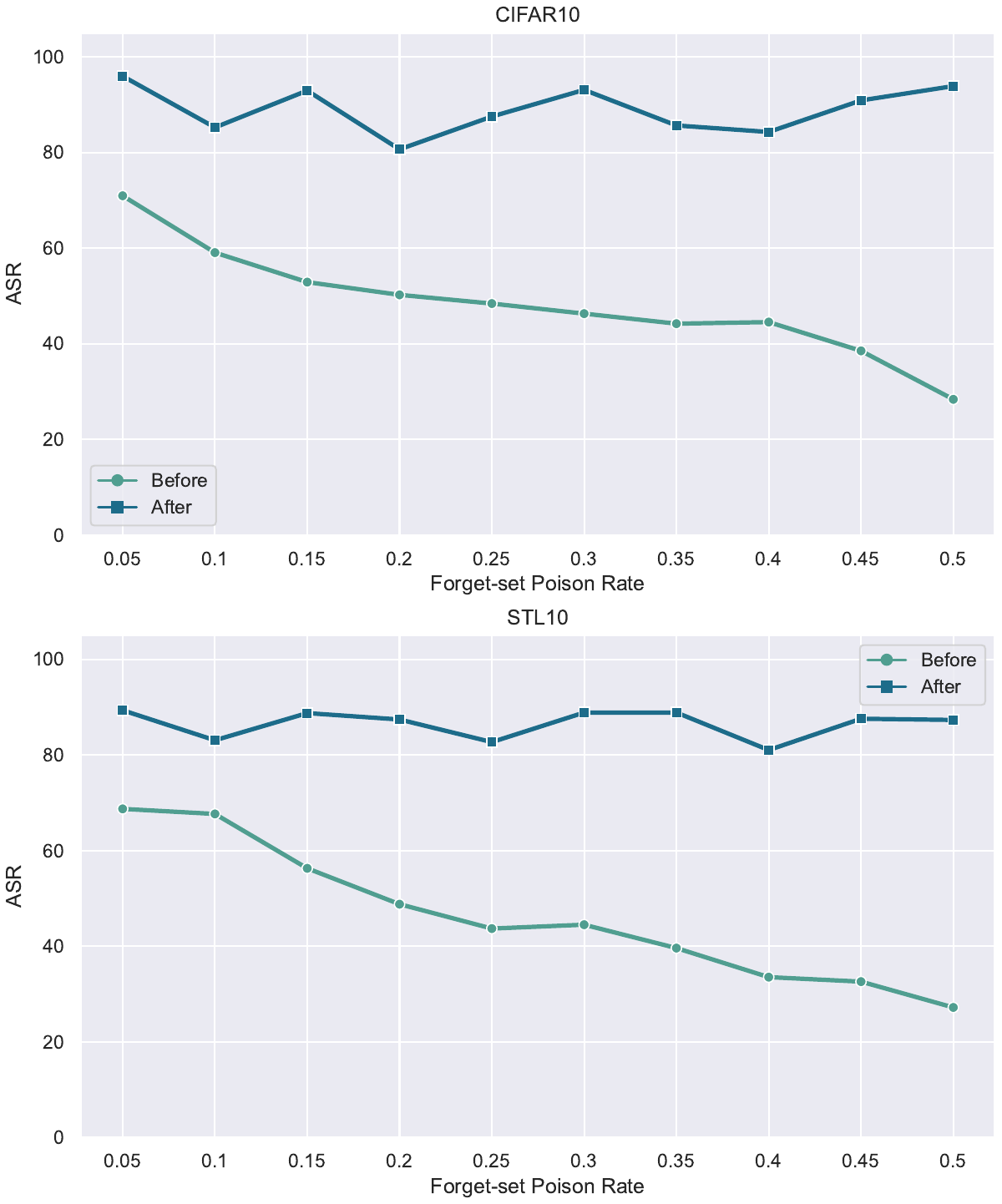}
    \caption{Evaluation of the attack changing the poison rate of the forget-set.}
    \label{fig:distantPercentage}
\end{figure}

In previous experiments, we established that poisoning data samples most distant from the target class effectively conceals the trigger signal; however, the proportion of such poisoned data also plays a critical role. In this additional experiment, we evaluate the backdoor success rate before and after unlearning with varying poisoning rates ranging from 0.05 to 0.50. The corresponding results are illustrated in Figure~\ref{fig:distantPercentage}. As expected, the ASR prior to unlearning is highest at the lowest poisoning rate and progressively decreases as the poisoning proportion increases. In contrast, the ASR post-unlearning remains relatively stable across all poisoning levels, with only minor fluctuations attributable to run-to-run variability.

\subsection{Evaluation of UNCLEAN against Defenses}
\label{sec:defensesEvaluation}

This section evaluates the resilience of the proposed attack against multiple defense mechanisms described in Section~\ref{sec:defenses}. In particular, we assess the extent to which the trigger can be embedded within the data and whether the resulting model parameters effectively obfuscate the attack by distributing the adversarial signal across classes. For this analysis, we focus on Fisher Forgetting, as it consistently demonstrates the highest vulnerability across the majority of evaluated datasets. The objective is to observe whether these defenses alter the behavior of the attack and, in turn, impact the model’s capacity to perform unlearning. The corresponding results are presented in Table~\ref{tab:CombinedDefenseEvaluation}.

\begin{table*}[!ht]
\tiny
\centering
\caption{Evaluation of the attack against three backdoor defenses: Neural Cleanse, Cognitive Distillation, and IBAU}
\label{tab:CombinedDefenseEvaluation}
\resizebox{0.85\textwidth}{!}{%
\begin{tabular}{c|ccc|ccc|ccc}
\toprule
\multirow{2}{*}{\textbf{Dataset}} & \multicolumn{3}{c|}{\textbf{Neural Cleanse}} & \multicolumn{3}{c|}{\textbf{Cognitive Distillation}} & \multicolumn{3}{c}{\textbf{IBAU}} \\
                                  & \textbf{Acc Retain} & \textbf{Acc Forget} & \textbf{ASR} & \textbf{Acc Retain} & \textbf{Acc Forget} & \textbf{ASR} & \textbf{Acc Retain} & \textbf{Acc Forget} & \textbf{ASR} \\ \midrule
MNIST                             & 98.7\%              & 0.0\%               & 68.1\%       & 99.4\%              & 0.0\%               & 89.1\%       & 99.2\%              & 0.1\%               & 71.0\%       \\
FashionMNIST                      & 84.5\%              & 0.0\%               & 99.8\%       & 89.8\%              & 0.0\%               & 85.6\%       & 93.6\%              & 0.0\%               & 58.1\%       \\
CIFAR10                           & 68.3\%              & 0.1\%               & 87.7\%       & 70.3\%              & 0.0\%               & 88.6\%       & 80.1\%              & 0.0\%               & 59.4\%       \\
STL10                             & 35.9\%              & 0.0\%               & 98.5\%       & 40.5\%              & 0.0\%               & 83.3\%       & 51.9\%              & 0.0\%               & 56.9\%       \\
\bottomrule
\end{tabular}%
}
\end{table*}

The results indicate that our attack remains largely resilient to defenses such as Neural Cleanse and Cognitive Distillation. In certain cases, the attack success rate even improves, while the model's ability to perform unlearning is preserved, as evidenced by a consistent 0\% accuracy on the forget-set. Notably, across all experimental runs, Neural Cleanse failed to identify the true backdoored class among the list of flagged suspicious classes, underscoring the stealthiness of the proposed method. Simultaneously, while Cognitive Distillation partially filters out contaminated samples, it does not substantially affect the final attack success rate, suggesting that the trigger is effectively concealed within the clean data distribution.

% The tables prove that our attack is only partly influenced by defenses such as Neural Cleanse and Cognitive Distillation. In certain scenarios, there is even an enhancement in the attack success rate, while preserving the model's unlearning capability, evidenced by a $0\%$ accuracy on the forget-set as anticipated. Notably, among the suspicious potentially backdoored classes identified by the Neural Cleanse defense, the actual backdoored class was never detected in any of the runs, showcasing the stealthiness of our strategy. At the same time, the Cognitive Distillation defense's filtering process effectively identifies only partially the contaminated data without limiting the final success rate, demonstrating that the generated trigger is sufficiently hidden within the clean data.

Defenses such as Neural Cleanse are ineffective in this context because their main assumption is that a backdoor trigger is exclusively and strongly correlated with a single target class. However, in our attack, the trigger is intentionally embedded within both the target class and a distant class, introducing conflicting gradients during optimization. Furthermore, the clean-label nature of the attack causes the trigger to become entangled with legitimate class features, thereby evading traditional backdoor detection mechanisms.

On the other hand, I-BAU is the only defense capable of significantly mitigating our attack, even if it does not fully prevent it. Unlike traditional methods, I-BAU does not attempt to explicitly identify the backdoor trigger. Instead, it searches for perturbations that the model is overly sensitive to and breaks the model’s reliance on them, regardless of their class origin. By doing so, I-BAU weakens the model’s ability to internalize the trigger before it can exploit it fully after forgetting.
Although I-BAU does not completely neutralize the backdoor, it reveals a promising avenue for advancing defense strategies.  It shows the importance of developing techniques that do not rely on class-specific information or assumptions about the trigger’s localization.

\subsection{Evaluation on Different Models}
\label{sec:differentModels}

\begin{figure}[!ht]
    \centering
    \includegraphics[width=\columnwidth]{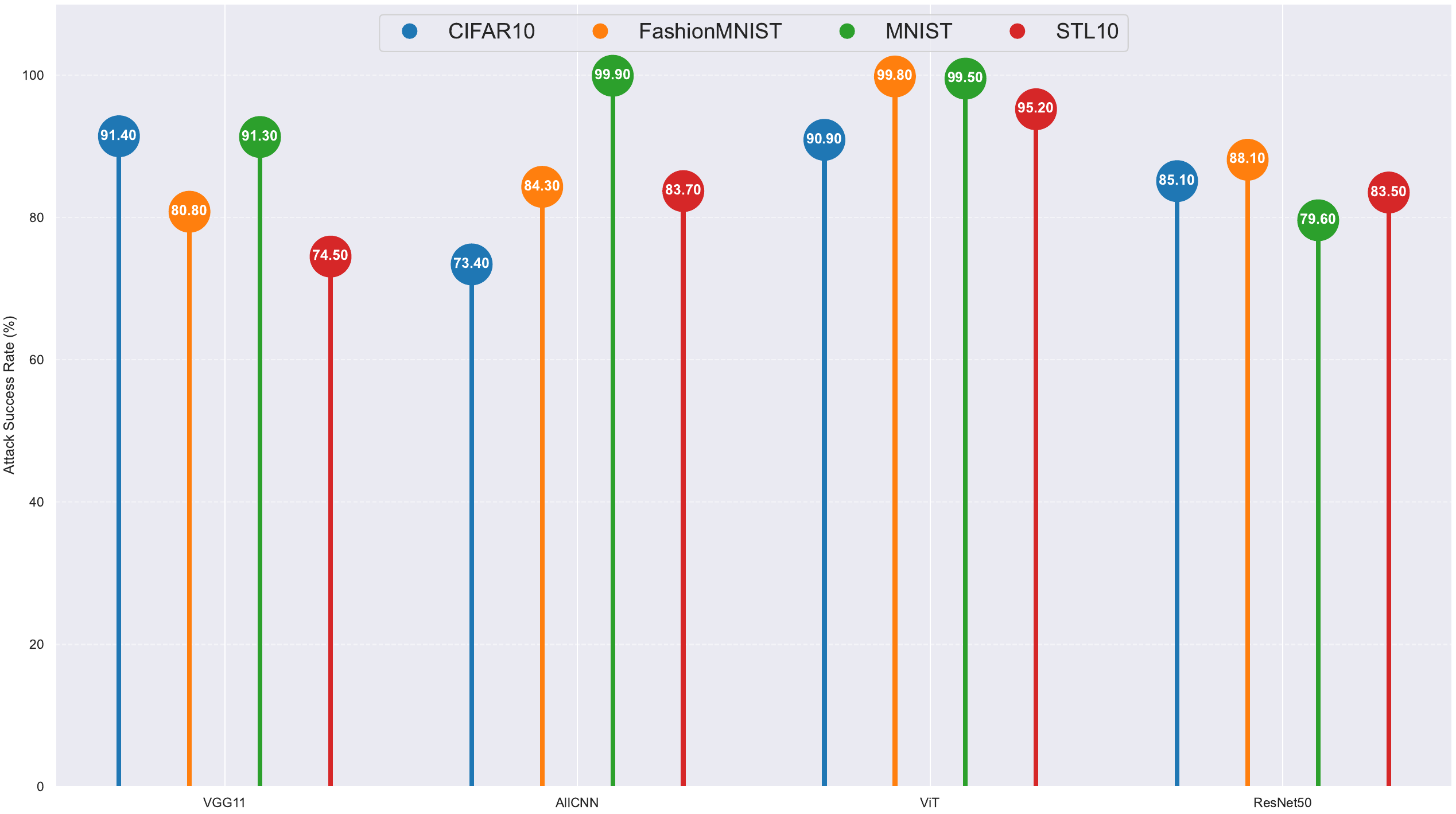}
    \caption{Evaluation of the attack on different model architectures.}
    \label{fig:models}
\end{figure}

This section focuses on evaluating our method against various models to determine the generalization of our attack. Specifically, in a realistic context, an attacker may lack knowledge of the models utilized by the system, potentially restricting their capacity to create effective triggers. 
Theoretically, our method aims to extract generalized features of the target class to produce a universal trigger applicable to different models trained for the same task, which we plan to test empirically.
To perform this experiment, we keep the previous baseline model, ResNet18, as the trigger generator and test against the models listed in Section~\ref{sec:datasetsModels}.
The results are reported in Figure~\ref{fig:models}.

% As we can see our attack effectively persists, even generating the trigger with a different model, achieving an attack success rate comparable to the baseline results.
% This confirms the generalizability of our approach even if the attacker has zero knowledge of the models used by the system, with state-of-the-art models as the ViT exhibiting highest ASR. ViTs, in contrast Lack strong inductive priors, so they tend to memorize and distribute features more globally.

Our results demonstrate that the proposed attack remains effective even when the trigger is generated using a different model, achieving an attack success rate comparable to the baseline. This highlights the general value of our approach, even under a zero-knowledge setting where the attacker lacks access to the victim model architecture. Notably, state-of-the-art models such as ViTs exhibit the highest attack success rates. Unlike CNNs, ViTs lack strong inductive biases, which leads them to memorize and distribute features globally. Their higher representational capacity enables them to make complex correlations, such as those introduced through clean-label poisoned triggers, without degrading performance on clean data. As a result, backdoor signals become deeply embedded within their internal representations, and unlearning procedures often fail to fully eliminate these influences, yielding higher post-unlearning attack success compared to lower-capacity models like CNNs.

\subsection{Comparison with Liu Attack}

As already anticipated, our attack lies between the strategies explained in ~\cite{huang2024uba} and ~\cite{liu2024backdoor}. In our attack, we explicitly differentiate from the first method as the forget-set utilized during unlearning comprises solely clean, unmodified data. In contrast, the ``attack with poison'' in~\cite{liu2024backdoor} strategy follows a method similar to ours. This section aims to compare the results achieved by the attack described in~\cite{liu2024backdoor} with those derived from our methodology. As the authors of the original paper do not provide an official code for replicating the results, we depend on the one presented in the article, employing an identical setting for our attack, in particular using the CIFAR10 dataset and $5\%$ of poisoning.
Results are presented in Table~\ref{tab:CompLiu}.

\begin{table}[]
\tiny
\centering
\caption{Comparison with attack with poisoning in~\cite{liu2024backdoor}}
\label{tab:CompLiu}
\begin{tabular}{ccccc}
\hline
\multirow{2}{*}{\textbf{Model Type}} & \multicolumn{2}{c}{\textbf{\begin{tabular}[c]{@{}c@{}}Attack with \\ Poisoning~\cite{liu2024backdoor}\end{tabular}}} & \multicolumn{2}{c}{\textbf{UCLEAN}}              \\ \cline{2-5} 
                                     & \textbf{$ASR_{before}$}                                       & \textbf{$ASR_{after}$}                                      & \textbf{$ASR_{before}$} & \textbf{$ASR_{after}$} \\ \hline
ResNet                               & 23.0\%                                                        & 45.0\%                                                      & 28.3\%                  & 83.3\%                 \\
VGG                                  & 23.0\%                                                        & 64.8\%                                                      & 28.3\%                  & 91.4\% \\  \hline          
\end{tabular}
\end{table}

As expected, in the attack described in ~\cite{liu2024backdoor}, data poisoning is confined to the target class while maintaining the original label without employing concealment strategies. This approach favors stealth but results in a trade-off with Attack Success Rate (ASR) after unlearning, thus yielding less than optimal outcomes. Our approach instead uses the Selection Strategy presented in Section~\ref{sec:selectionStrategy} to better conceal the trigger at training time, confirming the advantage introduced by our approach.

\subsection{Ablation Study}
\label{sec:abaltion}

This section focuses on an ablation study designed to evaluate the impact of various components of the proposed attack. As outlined in Section~\ref{sec:methodology}, the proposed method consists of three primary steps, and here we aim to determine whether our hypotheses are significantly influencing the attack. Specifically, this study compares the complete solution to versions where one or both of the initial steps, trigger generation and poison set selection, are omitted.
In particular, we tested four different scenarios: {\em (i)} Random Trigger/Random Selection, in which we remove our preparation strategies by utilizing a random trigger rather than an optimized one and by randomly choosing the forget set data for poisoning; {\em(ii)} Random Trigger, where we keep the proposed poison set selection but we use a trigger sampled from random noise; {\em (iii)} Random Selection, in which we employ the trigger generation but the poison set is sampled randomly; {\em (iv)} UNCLEAN, where we use the full solution.
In Table~\ref{tab:abalationStudy} we report the results of this ablation study assessing the changes in the ASR before and after unlearning according to the scenario.

\begin{table}[!ht]
\centering
\caption{ASR (\%) before and after unlearning across different attack scenarios. Random Trigger/Selection (RTS), Random Trigger (RT), Random Selection (RS) and Unclean}
\label{tab:abalationStudy}
\renewcommand{\arraystretch}{1.1}
\resizebox{\columnwidth}{!}{%
\begin{tabular}{l|cc|cc|cc|cc}
\toprule
\textbf{Scenario} & \multicolumn{2}{c|}{\textbf{MNIST}} & \multicolumn{2}{c|}{\textbf{FashionMNIST}} & \multicolumn{2}{c|}{\textbf{CIFAR10}} & \multicolumn{2}{c}{\textbf{STL10}} \\
 & ASR\textsubscript{before} & ASR\textsubscript{after} & ASR\textsubscript{before} & ASR\textsubscript{after} & ASR\textsubscript{before} & ASR\textsubscript{after} & ASR\textsubscript{before} & ASR\textsubscript{after} \\
\midrule
RTS & 100\% & 100\% & 100\% & 100\% & 89.5\% & 84.6\% & 75.6\% & 80.0\% \\
RT                   & 28.4\% & 74.8\% & 31.1\% & 89.9\% & 35.2\% & 85.8\% & 50.5\% & 90.2\% \\
RS                  & 100\%  & 100\%  & 99.7\% & 99.7\% & 86.3\% & 81.1\% & 90.5\% & 96.8\% \\
Unclean                           & 27.7\% & 88.2\% & 26.5\% & 93.4\% & 39.1\% & 92.7\% & 27.4\% & 91.7\% \\
\bottomrule
\end{tabular}%
}
\end{table}

As anticipated, removing the data selection criterion for poisoning the forget set has the most substantial impact on the performance of our attack. In this configuration, the attack success rate remains largely unchanged before and after unlearning. This outcome highlights the critical role of strategic data selection in enhancing both the stealth and effectiveness of backdoor attacks.

\begin{figure}[!ht]
    \centering
    \includegraphics[width=0.55\columnwidth]{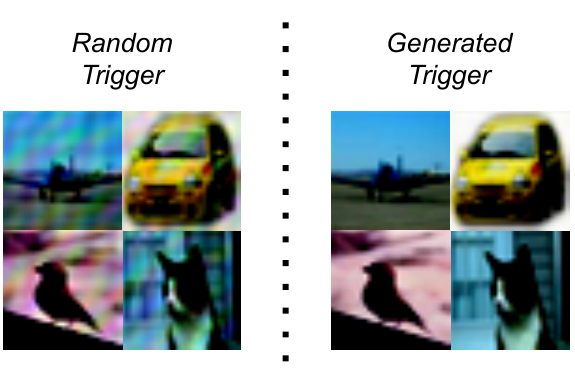}
    \caption{Comparison between random and generated trigger.}
    \label{fig:comparisonRandomOptimized}
\end{figure}

As shown in Figure~\ref{fig:comparisonRandomOptimized}, the random trigger visibly alters the image, making it even noticeable to the human eye. In contrast, our generated trigger preserves the original content, introducing minimal artifacts and rendering it almost imperceptible.

\section{Conclusion}
In this work, we presented UNCLEAN (UNlearning-activated CLEAN backdoor attack), a novel stealthy attack strategy that takes advantage of both the training and unlearning phases to compromise machine unlearning solutions. Following the lead of most advanced unlearning attacks, our approach maintains a fully clean unlearning phase by leveraging only non-poisoned samples. However, we overcome the current limitations of existing works by significantly expanding the attack success rate by injecting a non-targeted malicious signal distributed across multiple classes during the learning phase.
We evaluated UNCLEAN across multiple deep learning architectures and state-of-the-art unlearning techniques, showing its impact and robustness against existing defenses. Our findings highlight the effectiveness and severity of the proposed attack, with a remarkable improvement of over $32\%$ in attack success rate compared to previous similar methods.
Our results identify fundamental vulnerabilities in current machine unlearning solutions. This emphasizes the need for more resilient approaches and defense strategies to ensure the safe implementation of the ``Right to be Forgotten'' in the machine learning domain.


\section*{Acknowledgment}
This work was supported by the project ``GoTMaT - Governing Technology to Manage the Transition'' founded by the European Community - Next Generation EU, Mission 4 Component 2 Investment 1.3 - CUP B53C22003990006.


\end{document}